# Eye Disease Prediction using Ensemble Learning and Attention on OCT Scans

Gauri Naik, Nandini Narvekar, Dimple Agarwal, Nishita Nandanwar, Himangi Pande

Dr. Vishwanath Karad MIT WPU, Pune, Maharashtra, India

`gnaik826@gmail.com, nandininn194@gmail.com, agarwaldimple2406@gmail.com, nishitanandanwar@gmail.com, himangi.pande@mitwpu.edu.in`

**Abstract.** Eye diseases have posed significant challenges for decades, but advancements in technology have opened new avenues for their detection and treatment. Machine learning and deep learning algorithms have become instrumental in this domain, particularly when combined with Optical Coherent Technology (OCT) imaging. We propose a novel method for efficient detection of eye diseases from OCT images. Our technique enables the classification of patients into disease-free (normal eyes) or affected by specific conditions such as Choroidal Neovascularization (CNV), Diabetic Macular Edema (DME), or Drusen. In this work, we introduce an end-to-end web application that utilizes machine learning and deep learning techniques for efficient eye disease prediction. The application allows patients to submit their raw OCT scanned images, which undergo segmentation using a trained custom U-Net model. The segmented images are then fed into an ensemble model, comprising InceptionV3 and Xception networks, enhanced with a self-attention layer. This self-attention approach leverages the feature maps of individual models to achieve improved classification accuracy. The ensemble model's output is aggregated to predict and classify various eye diseases. Extensive experimentation and optimization have been conducted to ensure the application's efficiency and optimal performance. Our results demonstrate the effectiveness of the proposed approach in accurate eye disease prediction. The developed web application holds significant potential for early detection and timely intervention, thereby contributing to improved eye healthcare outcomes.

**Keywords:** Ensemble Learning, U-Net, Attention, OCT, CNV, DME, Drusen, InceptionV3, Xception

## 1 Introduction

The accurate detection and timely diagnosis of eye diseases play a crucial role in preserving vision and preventing further complications. Recent advancements in machine learning and deep learning algorithms, coupled with the use of Optical Coherence Tomography (OCT) imaging, have shown promising potential in this domain. OCT is a non-invasive imaging technique that provides high resolution, cross-sectional views of various eye structures, including the retina, macula, optic nerve, and choroid. By emitting low-intensity light beams and capturing reflections from different layers, OCT enables the identification of microscopic deformities, even in the early stages of diseases. Among the prevalent eye diseases, Choroidal Neovascularization (CNV) commonly occurs in association with age-related macular degeneration, while Diabetic Macular Edema (DME) affects a significant portion of the global diabetic population. Additionally, the deposition of yellow lipids and proteins known as DRUSEN can lead to various eye infections and diseases.

In this research, we propose a novel deep learning based model for the accurate prediction of these eye diseases using OCT images. Our methodology involves a two-step process: segmentation using a trained U-Net model, followed by disease classification using an ensemble model. The U-Net model, trained on a smaller subset of OCT images and corresponding manually annotated masks, performs segmentation on the rest of the raw OCT images. This leads to the production of a new dataset of segmented images that highlight specific regions of interest. These segmented images are then fed into the ensemble model, which combines the predictive capabilities of the Xception and Inception networks. To enhance the model's understanding of abnormalities within the extracted features, we introduce a novel self-attention layer at the final stage of each individual model training. Through the utilization of ensemble learning, our methodology adeptly harnesses and maximizes the intricate features within the OCT images, thereby enhancing the precision of disease classification. Notably, our approach surpasses the performance of current models in both accuracy and training time, despite being trained on a more compact yet diverse dataset.

To facilitate convenient and accessible eye disease detection, we have developed a user-friendly web application that allows individuals to upload their OCT images and receive prompt predictions through a graphical user interface (GUI). The users need to sign up and login to the portal to use it further and this ensures security of data provided by the users. All the data is stored in secured firebase servers in encrypted format for added level of security.

This research paper presents a comprehensive methodology, showcasing the working and effectiveness of machine learning, ensemble learning, and attention mechanisms in revolutionizing eye healthcare. It further speaks in detail about the dataset collection, preprocessing steps done on the dataset, website module and our results. Our approach aims to enable early detection and accurate prediction of eye diseases, contributing to improved patient outcomes and proactive healthcare strategies.

## 2   Related Work

This paper [1] introduced the Transformer model, which utilizes self-attention mechanisms for various natural language processing tasks, including machine translation. Although it primarily focuses on language tasks, the Transformer model has also been applied to image classification. In [2], authors presented the StandAlone Self-Attention (SASA) module, a lightweight self-attention mechanism designed specifically for image classification tasks. The authors demonstrated its effectiveness by incorporating it into various vision models, such as ResNet and MobileNet.

In [3], authors focused on disease recognition, achieving high accuracy using an eight layer AlexNet architecture. However, their evaluation lacked a comparison between AMD and DME images. In [4], authors addressed retinal cyst detection, implementing different CNN models for each cyst type. Despite achieving 79% accuracy, their approach could be further optimized and benefit from using an ensemble model. In [5], authors proposed a deep learning-based approach for segmenting retinal fluid in OCT images, achieving state-of-the-art performance using ensemble models. In [6], authors focused on classifying macular OCT images, utilizing a multi-scale CNN ensemble approach and achieving an accuracy of 95.9%. In [7], authors developed an architecture combining deep residual networks, recurrent neural networks, and inception modules for multi-class retinal disease classification. In [8], authors proposed an ensemble model called OCTx for classifying and detecting eye diseases using OCT images. However, their model was trained only on a specific dataset, which may limit its robustness. In [9], authors compared various pre-trained models for classifying eye diseases, with DenseNet169 achieving the highest accuracy. However, no ensemble of models was used. In [10], authors focused on classifying macular edema using ResNet, achieving high accuracies but with limited dataset robustness. In [11], authors proposed an ensemble model for computer-aided diagnosis of retinal diseases, employing multi-resolution features and achieving accuracy in detecting six retinal diseases. In [12], authors utilized deep learning and ensemble learning methods for classifying retinal diseases, achieving improved performance with CNN base classifiers and ensemble models. In [14], authors addressed the detection of choroidal neovascularization (CNV) using OCT images, employing a machine learning scheme and achieving promising results. However, their research gap lies in the lack of comprehensive evaluation on different classifiers and feature optimization techniques.

Overall, while these studies have made advancements in eye disease detection and classification, there are opportunities for further improvements and addressing research gaps, such as better evaluation methods, robustness with diverse datasets, and ensemble model utilization.

## 3   Research Methodology

In this section, we outline the methodology employed for the development of our eye disease prediction web application. Fig. 1 shows the system architecture of our proposed model. The methodology consists of the following steps:

### 3.1   Data Collection and Preprocessing

The data used in this research was obtained from a publicly available dataset provided by Kermany et al. in 2018 [16]. The dataset comprises a total of 207,103 OCT images divided into four classes: Choroidal Neovascularization (CNV), Diabetic Macular Edema (DME), DRUSEN, and Normal eyes. Figure 2 showcases sample images from the dataset. Specifically, the dataset consists of 108,312 training images, with 37,206 images representing CNV, 11,349 images representing DME, 8,617 images representing DRUSEN, and 51,140 images representing normal eyes. To evaluate the performance of the model, we used 1,000 test images, with an equal distribution of 250 images for each category.
Fig. 2 shows sample of the various classes of images included in the dataset.

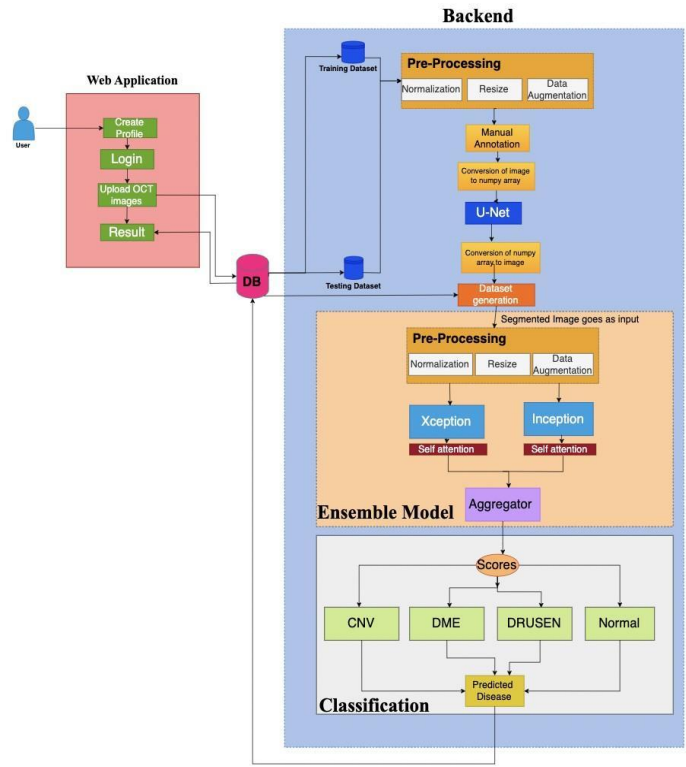

**Fig.1.** System Architecture

Since the publicly enabled dataset consists solely of raw OCT images belonging to the 4 classes of diseases, there was no existing source of ground-truth images, better known as masks. In order to fill this gap, we took 8 images per class and manually annotated these images for demarcating the disease affected regions. Together, these 32 images acted as masks for the raw images that together formed the input set for our custom U-Net model. The annotation process involved marking the hyper-reflective zones and affected layers within the inner retinal layer. These annotations served as the masks for the U-Net model training input. Fig. 3 illustrates an example of an annotated image and its corresponding mask, depicting a patient affected by Diabetic Macular Edema (DME).

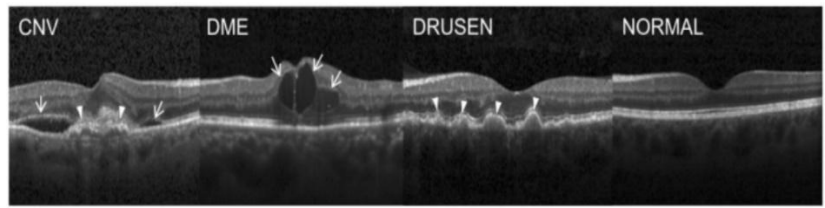

**Fig.2.** OCT Images with Eye Diseases

The careful collection and preprocessing of the dataset, along with the manual annotation of key regions, lay the foundation for subsequent stages of the research methodology.

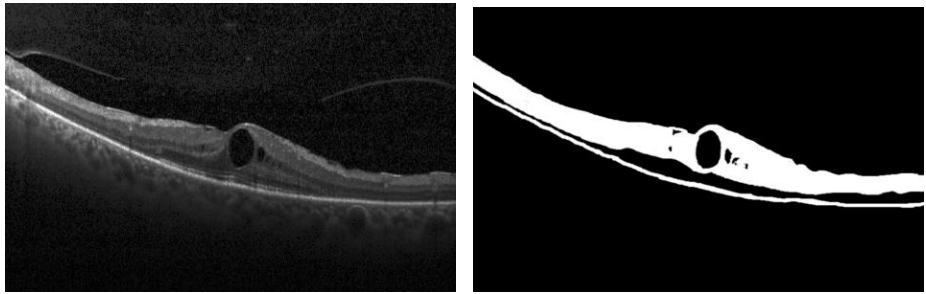

**Fig.3.** OCT Scan image (Left) and mask of the scan (Right) of patient affected with DME

To train the U-Net model and generate the dataset for training and preprocessing, the OCT images and their corresponding masks were converted into numpy arrays. These arrays were then used to train our neural network for the task of doing the binary semantic segmentation of the images. The arrays are split into 80-20 split; 80% of the data acts as the training set and 20% acts as the testing set. Training the U-Net model with these 32 images enabled the segmentation of the entire dataset, accurately delineating the affected regions in the OCT images and creating a more precise dataset. The masks and segmented images were instrumental in defining the thickness of the hyper-reflective layers and the total thickness of the inner retinal layers, as well as identifying abnormalities for successful disease classification.

Due to computational constraints on the hardware used for model training, we randomly selected 500 images from each class to train the classification models. These images were utilized to train the models for accurate disease classification.

In the following sections, we describe the model architecture, training process, and evaluation metrics used for accurate eye disease prediction.

For each of the 32 cases we mark the hyper reflective zones (thin lines on the lower side as shown in Fig. 3 and 5) and the affected layers in the inner retinal layer. These annotations act as the masks of images that will be used as the input for our U-Net.

In our study, we employed preprocessing techniques on both segmented and raw OCT images to enhance the quality of the data and improve the performance of our model. For both segmented and raw OCT images, we applied several preprocessing steps to standardize the data and ensure compatibility with our classification model. These steps included normalizing, resizing. Normalization was performed to adjust the pixel values of the images and bring them to a standardized scale. By normalizing the images, we aimed to reduce the impact of variations in pixel intensity across different images, thus enabling the model to focus on relevant features rather than absolute pixel values. Resizing was another essential preprocessing step, which involved resizing all images to a consistent resolution. This step was necessary to ensure that the input images have a uniform size, facilitating the model to learn and extract meaningful features effectively. In our study, The input images to the U-Net model were resized to 496 x 496 and the segmented images were resized to a resolution of 299 x299 pixels, which was found to be optimal for inception and xception model. Overall, the preprocessing of both segmented and raw OCT images involved normalizing the pixel values, resizing the images to a consistent resolution. These steps were crucial in preparing the data for subsequent analysis and classification using our ensemble model.

### 3.2   U-Net Model for Segmentation

An OCT scan gives extremely high resolution cross-sectional scans of the eye retina owing to which the various layers of retina are very precisely visible in a good quality OCT image. It is a completely non-invasive procedure during which a number of light waves strike the retinal tissues, and the reflected light is used for creating a detailed three-dimensional(3D) image of the eye. All the inner retinal layers of the retina have different refractive indices and hence once light gets reflected from it; the layers show up in the scan in different shades. In such an OCT scan that is clearly and precisely captured, one can see the multitude of retinal layers that are present inside the eye as shown in the below image.

Most OCT images have salt and pepper noise, and each image has a different distribution. It is difficult to distinguish background and retina layers using a single threshold. Therefore, the pre-processed images pass through the U-Net architecture to remove the noise and segment the image into the affected and not affected region. For the implementation of our project, we resize all images to 496 x 496 size and then, we segment all these images so that irrespective of the disease, the neural network is correctly able to classify the affected layers in the retina.

A U-Net is a type of convolutional neural network architecture used for image segmentation. The name "U-Net" comes from the network's U shape. The network consists of two parts: the contracting path i.e., the down sampling and the expansive path i.e., the upsampling.

The contracting path or downsampling (also known as pooling) is a traditional convolutional neural network, which consists of several convolutional and max pooling layers. This path is used to capture the context and features of the input image. This is a process where the spatial resolution of an image is reduced while retaining the most important features. This is typically done using a convolutional layer with a large stride. The down sampling operation is applied several times in the encoding path of the U-Net architecture, which reduces the size of the input image but captures the most important features.

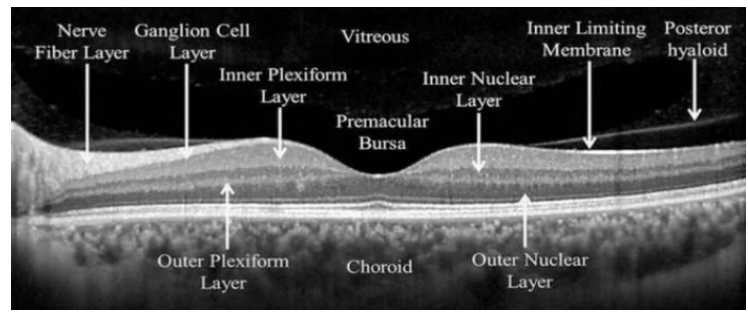

**Fig.4.** OCT scan showing different retinal layers

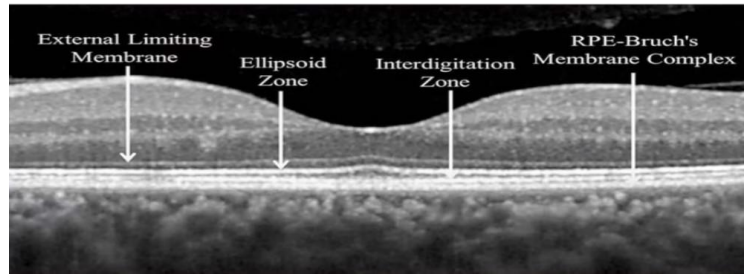

**Fig.5.** OCT scan showing different layers in hyper-reflective zone of retina

The expansive path or upsampling (also known as deconvolution or transposed convolution) is used to enable precise localization and segmentation of the image. It is composed of up-convolutional layers and skip connections that copy feature maps from the contracting path and concatenate them with feature maps from the up-convolutional layers. This allows the network to recover spatial information that may have been lost during the down-sampling operation of the contracting path. The upsampling operation is done by transposed convolution, where the output tensor is padded with zeros and then convolved with a filter. This results in a larger feature map with the same number of channels as the original image. Additionally, skip connections are used to concatenate feature maps from the encoder with those from the decoder, which allows the decoder to make use of low-level features captured in the encoder. By using a combination of downsampling and upsampling, U-Net is able to learn a hierarchical representation of the input image and capture fine details while still retaining contextual information. Fig. 6 showcases the architecture of the U-Net Model used in our proposed method.

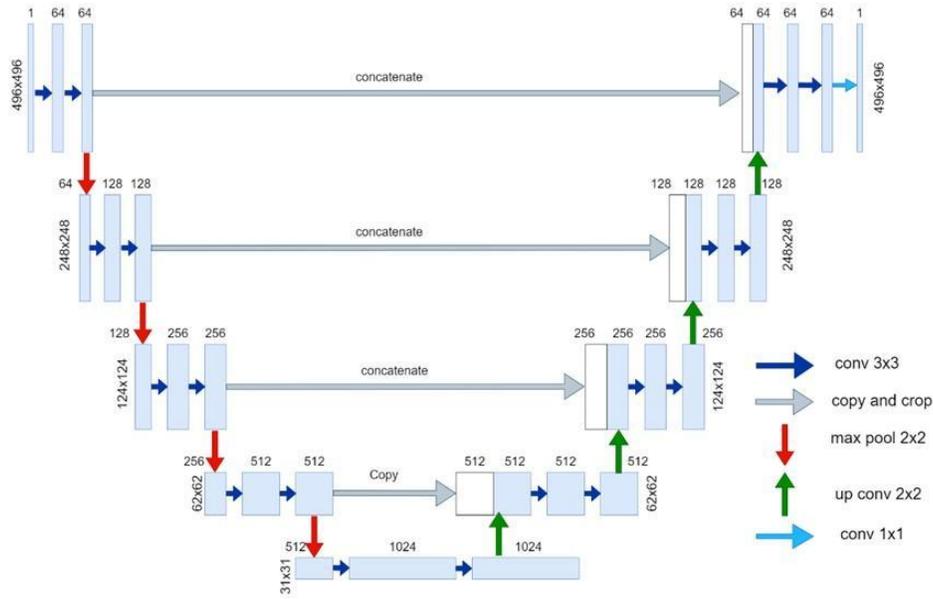

**Fig.6.** U-Net Architecture in proposed solution

### 3.3 Self-Attention Mechanism

Attention mechanisms play a crucial role in enhancing the model's ability to accurately predict abnormalities in OCT images. By assigning weights to different parts of the input image, attention mechanisms allow the model to focus on the most discriminative regions while downplaying less relevant areas.

To incorporate self-attention mechanisms into our image classification algorithm, we utilized feature maps obtained from the pretrained model (Inception and Xception). The feature maps served as the input for calculating attention weights based on queries, keys, and values. The queries, keys, and values were derived through linear transformations ($W_g$, $W_f$, and $W_h$) applied to the feature maps.

$$Key : f(x) = \mathbf{W}_f x \quad (1)$$
$$Query : f(x) = \mathbf{W}_g x \quad (2)$$
$$Value : f(x) = \mathbf{W}_h x \quad (3)$$

$$\alpha_{i,j} = softmax(f(x_i)^T g(x_j)) \quad (4)$$

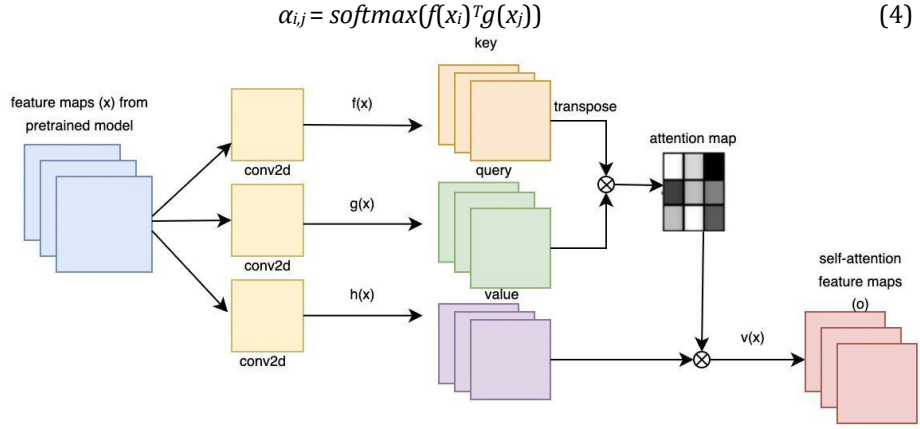

**Fig.7.** Self-Attention Layer

The attention weights (4) were then computed using the dot-product attention mechanism, which measures the similarity between different positions within the feature maps. The formula for calculating the attention weights is as follows:

Here, $x_i$ and $x_j$ represent different positions in the feature maps. Once the attention weights are obtained, they are applied to the values (3) to generate the attended feature maps. The attended feature maps highlighted the most relevant information within the OCT image, incorporating both local and global contextual cues. This integration was achieved through element-wise addition between the attended feature maps and the original feature maps.

$$o_{i,j} = W_v\left(\sum_{i=1}^{N} \alpha_{i,j} h(x_i)\right) \quad (5)$$

The final output of the self-attention mechanism (5) represented the self attention feature maps. These feature maps encapsulated the selectively focused and enhanced representation of the OCT image, capturing intricate patterns and subtle abnormalities.

By employing the self-attention mechanism in our OCT image classification system, we significantly improved its discriminative power and ability to accurately classify images. The attention-based approach allowed the model to effectively capture the relationships between different positions within the feature maps, resulting in a more robust and accurate classification system.

### 3.4 Ensemble Model Construction

To enhance the accuracy and robustness of our disease classification system, we adopt an ensemble model that combines multiple individual models. We begin by training several well-established models individually, such as ResNet-50, InceptionV3, Xception, VGG16, and DenseNet, utilizing pre-trained weights from the ImageNet dataset. During the individual training process, we incorporate self attention mechanisms on the feature maps from these pretrained models. This integration enables the models to selectively emphasize abnormalities and focus on disease-specific features, thereby improving their classification performance.

Throughout the training phase, we carefully optimize the model parameters for each individual model, ensuring that they learn to allocate attention effectively. By incorporating self-attention, the models dynamically assign higher weights to regions of interest, enhancing their ability to accurately classify diseases in segmented images. This process contributes to the overall discriminative power and robustness of the models.

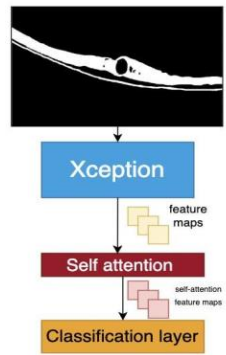

**Fig.8.** Individual model

After training, we conduct a comprehensive evaluation of each individual model's performance. Based on accuracy, we identify the top two performing models, Xception and Inception, with accuracies of 86% and 90%, respectively. These models exhibit exceptional classification capabilities and demonstrate superior performance on the given disease classification task.

To construct our ensemble model, we combine the predictions generated by the selected individual models. By considering the collective decisions of these models, the ensemble model can effectively mitigate individual biases and uncertainties, leading to more reliable and accurate disease classification outcomes. The combination of multiple models in the ensemble approach leverages their diverse strengths and perspectives, resulting in improved overall accuracy and robustness.

### 3.5    Web-Application Module

In the development of our disease classification system, we implemented a web application module using various technologies such as Flask, Firebase, HTML, and CSS. The purpose of this module is to provide users with a user-friendly interface to interact with the system and obtain disease classification results for OCT images.

The web application begins with a user registration process, where users create their profiles by providing necessary information and credentials. Once registered, users can log in to the web application using their credentials to access its features. Upon logging in, users are presented with a visually appealing front-end design characterized by a blue and white color scheme. The aesthetic layout includes pages such as the home page, about us, contact us, and other relevant sections to enhance user experience.

To perform disease classification, users are required to upload OCT images captured through their device's camera. The uploaded OCT image is then processed through a U-Net architecture and an ensemble model. The U-Net architecture is responsible for segmenting the OCT image and extracting disease-specific features, while the ensemble model combines multiple well-established models, including DenseNet201, ResNet, and Inception, to make accurate disease classification predictions. As the user uploads an OCT image, the system saves it in the database. Finally, the web application authenticates the disease classification results to the user, providing the predicted disease category or any relevant information through the website interface. This enables users to easily access and interpret the classification outcomes. By integrating user registration, OCT image uploading, ensemble model processing, and result authentication, the web application provides a seamless and efficient experience for users seeking accurate disease classification results for OCT images.

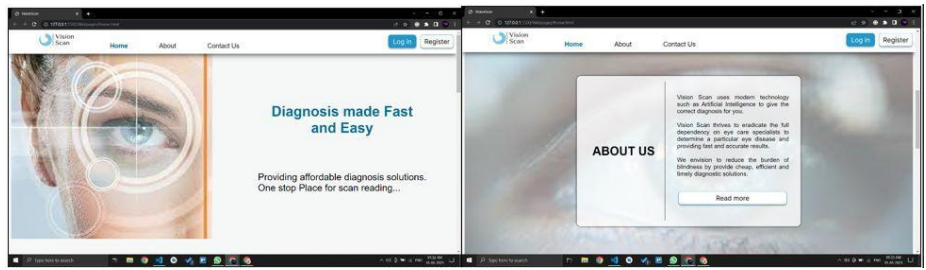

**Fig.9.** Website UI

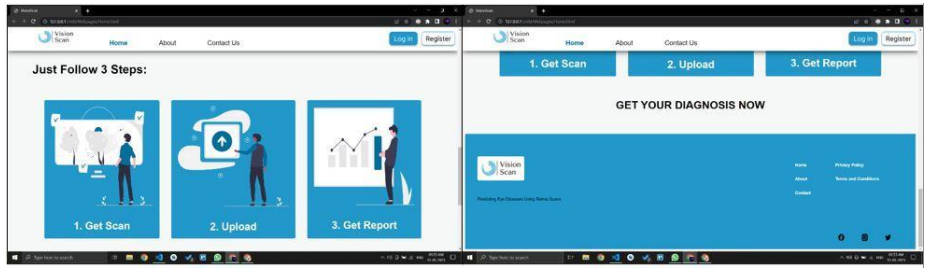

**Fig.10.** Website Home Page UI

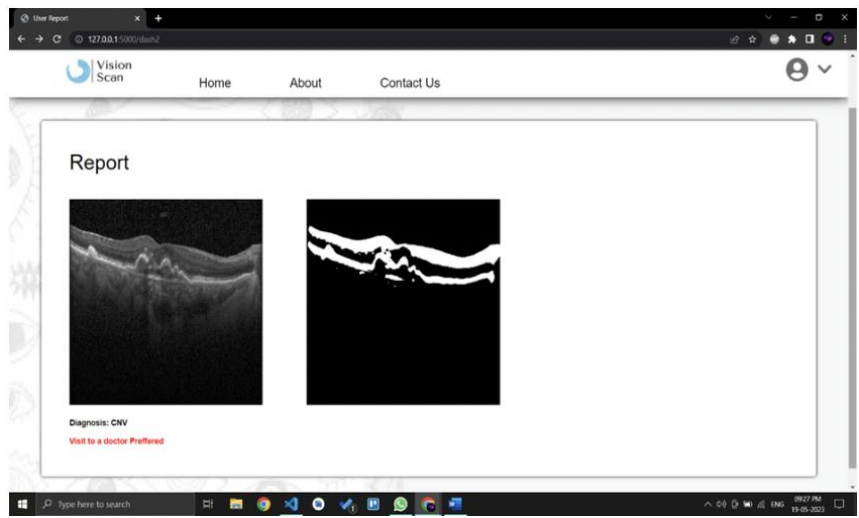

**Fig.11.** Report generated by Website

## 4    Result and Analysis

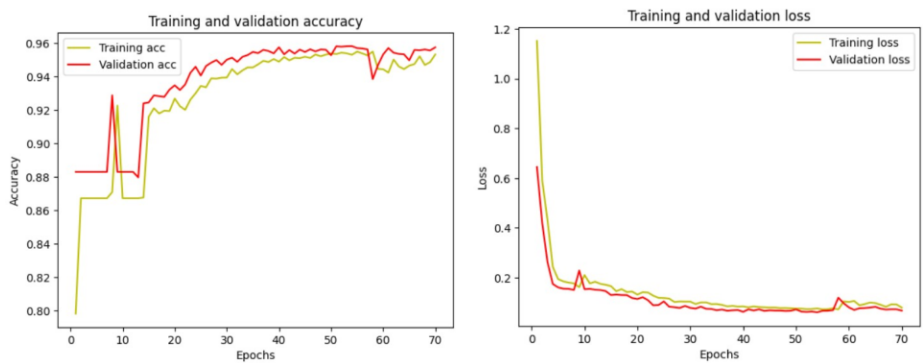
**Fig.12.** Accuracy and Loss Metrics of U-NET model

The Fig. 12 represents the loss and accuracy of the U-Net Model. The U-Net Model gives an accuracy of **94.69**% on the training dataset with 32 images and their respective masks as target image. The U-Net model was trained for 70 epochs and it gave IOU (Intersection Over Union) of **84.078**%. Increasing the epochs number or the number of images in the training set leads to overfitting of the model which points out the fact that this is an optimum model at the current position.

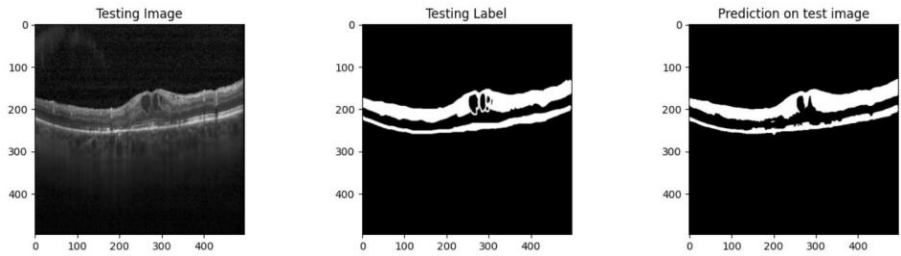

**Fig.13.** Analysis of image after passing through custom U-Net architecture

Each of the individual models was added with Self-attention layers to classify the image into disease categories. Each such model was run with 2000 images in the training set and 968 images in the testing set (distributed evenly across the 4 classes) in order to get a generalized accuracy.

**Table 1.** Individual model accuracies

| Model | Accuracy |
| --- | --- |
| Xception | 90% |
| InceptionV3 | 86% |
| DenseNet201 | 84% |
| ResNet50 | 58% |
| Vgg16 | 55% |

Table 1 showcases the accuracy of each model on the same testing set to give the general view. We take the 2 models with highest accuracy, i.e. Xception and InceptionV3 to create the ensemble model.

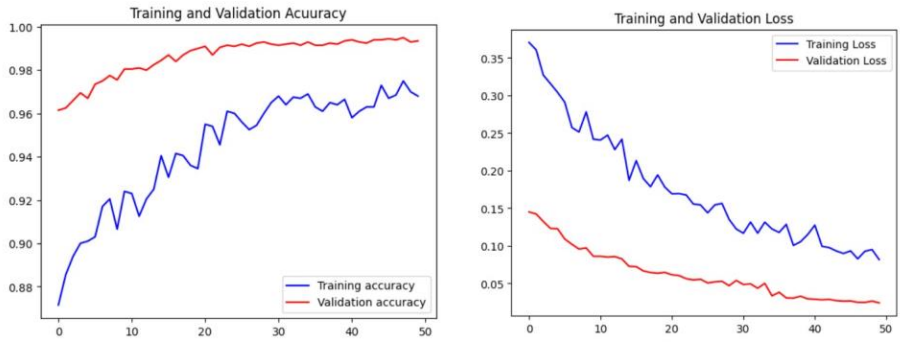

**Fig.14.** Accuracy and Loss Metrics of ensemble model across 50 epochs

This final model gives a cumulative accuracy of **96.69**% which is significantly higher than the individual model accuracies. Other evaluation metrics for the ensemble model are (Table 2):

**Table 2.** Ensemble Model Evaluation Metrics

| Metric | Value |
| --- | --- |
| Accuracy | 0.96694% |
| Precision | 0.96708% |
| Recall | 0.96694% |
| F1-score | 0.96690% |

Table 3 showcases comparative analysis between our proposed model and other models that used ensemble learning and other CNN models for eye disease prediction using OCT images.

**Table 3.** Comparison with other models

| Method | Accuracy |
| --- | --- |
| Singh et al. , 2021 [18] | 0.914% |
| Pin et al., 2021 [17] | 0.916% |
| Lu et al., 2018[15] | 0.959% |
| Our Proposed model | 0.966% |

## 5   Conclusion

Eye diseases are one of the major concerns around the globe. Thus, a lot of time and effort is dedicated to the early detection, prevention and treatment of eye diseases. As technology advances, utilizing AI to help solve these problems is common. For this purpose, the OCT imaging technique is widely used for its non-invasive nature and for getting a detailed cross-section view of the retina (a part of the eye).

The proposed model uses the ensemble learning approach and is integrated into a user-friendly web application to democratize access. This step also paves the way for future adaptations to include additional diseases.

Due to hardware limitations, the current model was trained on a subset of data; however, future improvements involve training the ensemble model on the complete dataset, enhancing its capabilities. Advanced techniques such as GANs and attention mechanisms show promise in enhancing anomaly detection and classification accuracy. The integration of GANs can aid in generating synthetic data to mitigate data scarcity challenges. Collaboration with medical professionals will further validate and refine the system, ensuring alignment with practical medical applications.